\documentclass[traditabstract]{aa}  
\usepackage{graphicx,psfig,longtable,lscape,amssym,natbib,color}  
\newcommand{\ltsima} {$\; \buildrel < \over \sim \;$}  
\newcommand{\gtsima} {$\; \buildrel > \over \sim \;$}  
\newcommand{\lta} {\lower.5ex\hbox{\ltsima}}  
\newcommand{\gta} {\lower.5ex\hbox{\gtsima}}  
\newcommand{\Ha} {H$\alpha$}  
\newcommand{\Hb} {H$\beta$}  

\newcommand{\ergsHz}{\ensuremath{\mathrm{\,erg}{\mathrm{\,s}^{-1}{\mathrm{\,Hz}^{-1}}}}}
\newcommand{\ergs}{\ensuremath{\mathrm{\,erg}{\mathrm{\,s}^{-1}}}}

\newcommand{\kms}{$\rm{\,km \,s}^{-1}$}
\newcommand{\forb}[2]{\mbox{$[{\rm #1\, #2}]$}}
\newcommand{\oiii}{\forb{O}{III}}

\begin{document}

\title{Exploring the spectroscopic properties of relic
  radiogalaxies.\thanks{Based on observations made with the Italian
    Telescopio Nazionale Galileo operated on the island of La Palma by the
    Centro Galileo Galilei of INAF (Istituto Nazionale di Astrofisica) at the
    Spanish Observatorio del Roque del los Muchachos of the Instituto de
    Astrof´ısica de Canarias.}}

\titlerunning{Spectroscopic properties of relic radiogalaxies}
\authorrunning{Capetti et al.}
  
\author{Alessandro Capetti \inst{1} \and Andrew Robinson \inst{2} \and Ranieri
  D. Baldi \inst{3} \and Sara Buttiglione \inst{4} \and David J. Axon
  \inst{2,5} \and Annalisa Celotti \inst{3,6} \and Marco Chiaberge \inst{7,8} }
   
\offprints{A. Capetti}
     
\institute{INAF - Osservatorio Astrofisico di Torino, Strada Osservatorio 20,
  I-10025 Pino Torinese, Italy \and Department of Physics, Rochester Institute
  of Technology, 85 Lomb Memorial Drive, Rochester, NY 14623 \and SISSA-ISAS,
  Via Bonomea 265, 34136 Trieste, Italy \and INAF, Osservatorio Astronomico di
  Padova, Vicolo dell'Osservatorio 5, I-35122 Padova, Italy \and School of
  Mathematical and Physical Sciences, University of Sussex, Falmer, Brighton
  BN1 9RH, UK \and INAF - Osservatorio Astronomico di Brera, Via E. Bianchi
  46, I-23807 Merate, Italy \and Space Telescope Science Institute, 3700 San
  Martin Drive, Baltimore, MD 21218, U.S.A. \and INAF-Istituto di Radio
  Astronomia, via P. Gobetti 101, I-40129 Bologna, Italy} \date{}

\abstract{From an optical spectroscopic survey of 3CR radiogalaxies (RGs) with
  z$<$0.3, we discovered three objects characterized by an extremely low level
  of gas excitation and a large deficit of line emission with respect to RGs
  of similar radio luminosity. We interpreted these objects as relic active
  galactic nuclei (AGN), i.e., sources observed after a large drop in their
  nuclear activity.

  We here present new spectroscopic observations for these three galaxies and
  for a group of ``candidate'' relics. None of the candidates can be
  convincingly confirmed.

  From the new data for the three relics, we estimate the density of the
  line-emitting gas. This enables us to explore the temporal evolution of the
  line ratios after the AGN ``death''. The characteristic timescale is the
  light-crossing time of the emission line region, a few $\sim$$10^3$ years,
  too short to correspond to a substantial population of relic RGs. Additional
  mechanisms of gas ionization, such as ``relic shocks'' from their past high
  power phase or stellar sources, should also be considered to account for the
  spectroscopic properties of the relic RGs.

  Relic RGs appear to be a mixed bag of sources in different phases of
  evolution, including AGN recently ($\sim$$10^4$ years ago) quenched,
  galaxies that have been inactive for at least $\sim$$10^6$ years, and
  objects caught during the transition from a powerful RG to a low power
  FR~I source.
 
}
\maketitle
  
\section{Introduction}
\label{introduction}

With the aim of gaining a better understanding of the properties of the
central engines of RGs, we recently performed an optical
spectroscopic survey of the 113 RGs belonging to the 3CR sample and
with $z<$ 0.3 \citep{spinrad85}, using the Telescopio Nazionale Galileo
\citep{buttiglione09,buttiglione11}. The 3CR sources show a bimodal
distribution of excitation index, a spectroscopic indicator that measures the
relative intensity of low- and high-excitation lines. This unveils the presence
of two main sub-populations of radio-loud AGN, high and low excitation
galaxies (HEGs and LEGs, respectively) \citep{buttiglione10}. We speculated that
the differences between LEGs and HEGs are related to a different mode of
accretion.  

In addition to the two main classes, we find three galaxies (namely, 3C~028,
3C~314.1, and 3C~348) that stand out for their low [O~III]/\Hb\ ratio of
$\sim$0.5. For this reason, in \citet{buttiglione09} we refer to this class as
extremely low excitation galaxies (ELEGs). In the optical spectroscopic
diagnostic diagrams, they are well separated from the rest of the 3CR sample
and fall in a region scarcely populated by emission-line sources from the
Sloan Digital Sky Survey (SDSS, \citealt{kewley06}). They are located,
considering their [S II]/\Ha\ and [O I]/\Ha\ ratios, well within the AGN
region. In the [O III]/\Hb\ vs [N II]/\Ha\ plane they straddle the line
separating AGN from the so-called composite galaxies.

These objects also stand out for their radio properties: 3C~028 and 3C~314.1
do not show the presence of a radio core at a level of 0.2 and 1.0 mJy,
respectively (see \citealt{giovannini88} and references therein), while a 10
mJy radio core is detected in 3C~348 \citep{morganti93}. The core dominance,
i.e., the ratio between the core (at 5 GHz) and total (at 178 MHz) radio
emission, for these three sources is in the range F$_{\rm core}$/F$_{\rm tot}
\lesssim 10^{-5} - 10^{-4}$ compared to an average ratio for RGs of similar
luminosity of F$_{\rm core}$/F$_{\rm tot} \sim10^{-2.8}$.  Similarly, they
have a [O III] line luminosity that is lower by 1 - 2 orders of magnitude with
respect to RGs at similar radio luminosity.

Based on these findings, we interpreted these objects as {\sl relic} AGN in
which nuclear activity is currently switched off (or strongly reduced with
respect to its long term average level, \citealt{relic}). The considered
quantities (radio core, emission-line luminosities and ratios, and total radio
emission) respond to changes in the AGN activity level on different
timescales. In fact, they originate from scales ranging from sub-parsec for the
radio core, to kpc for the line emission produced in the narrow line region
(NLR), and to hundreds of kpc for the extended radio emission.

Thus, a drop in the ionizing photon luminosity would result in recombination
and cooling of the NLR, leading to a relatively low level of [O~III] emission
and to a decrease of [O~III]/H$\beta$ ratio, while the extended radio emission
remains essentially unchanged.

All relevant timescales of the spectral evolution of the NLR depend on the
inverse of the gas density, but, from our survey spectra we unfortunately
could not obtain an accurate estimate of this quantity. In fact, our spectrum
of 3C~348 does not cover the [S~II] doublet, while in 3C~028 the
[S~II]$\lambda$6731 line falls in a CCD defect. For 3C~314.1 we obtain a ratio
[S II]$\lambda$6716/[S II]$\lambda$6731 = 1.6 $\pm$ 0.3, which is only
indicative of a low density ($n_e \lesssim 5\times 10^2$) regime.

In the 3CR sample there are also seven galaxies that can be considered as
``candidate relics''.  They could not be classified spectroscopically (since at
least one of the key emission-line could not be detected); however, since they
show a low \oiii\ luminosity with respect to their total radio luminosity, we
consider them as candidate relics.

Further spectroscopic data are then needed to i) improve the census of relic
RGs in the 3CR sample and ii) measure their NLR gas density. The aim of this
paper is to present new data for the relics and the candidate relics so as to
further explore their spectroscopic properties and gain a better
understanding of the nature of these sources.  The paper is organized as
follows: in Sect. \ref{observations} and \ref{sect2} we present the new
observations and provide the fluxes of the key emission-lines. In the
following sections we explore the time evolution of line luminosities and
ratios after a drop in the nuclear intensity (Sect. \ref{sect3}), the relation
between relics and active galaxies (Sect. \ref{lifetime}), and the
alternative scenarios for the nature of the relic RGs (Sect. \ref{origin}). A summary is given in Sect. \ref{summary}.

Throughout, we use $H_{\rm o} = 71$ km s$^{-1}$ Mpc$^{-1}$, $\Omega_{\Lambda}
= 0.73$ and $\Omega_{\rm m} = 0.27$.

\begin{table}
  \begin{center}
\caption[Journal of the observations.]{Journal of the observations.} 
\label{logoss} 
    \begin{tabular}{l | c | c | c  c| c c c}
      \hline \hline
Name & z & Date  & \multicolumn{2}{|c|}{ VHR-V }
&\multicolumn{3}{|c}{Red grism} \\
&              &             & n   & T$_{\rm exp}$ &    & n    &  T$_{\rm exp}$  \\ \hline 
3C~028     & 0.1953  & 29Jul09    &  &  & HRI & 3 & 800  \\    
3C~314.1   & 0.1197  & 14Apr09    &  &  & HRR & 4 & 800  \\    
3C~348     & 0.1540  & 03Mar09    &  &  & HRI & 4 & 800  \\    
\hline
3C~035     & 0.0670  & 23Sep09    & 2 & 750 & HRR & 2 & 750  \\    
3C~035     & 0.0670  & 23Sep09    & \multicolumn{5}{l}{2 \,\,\,\,\,\,\,\,750\,\,\,\,  (both with LRB)}\\
3C~089     & 0.1386  & 02Feb09    & 2 & 800 & HRR & 2 & 800  \\    
3C~173.1   & 0.2921  & 15Mar09    & 2 & 800 & HRI & 3 & 800  \\    
3C~258     & 0.1650  & 29Mar09    & 2 & 800 & HRI & 3 & 800  \\    
3C~438     & 0.2900  & 21Jul09    & 2 & 800 & HRI & 3 & 800  \\    
\hline
\end{tabular}                                               
  \end{center}                                                  
  Column description: (1) name of the source; (2) redshift; 
  (3) night of observation;
  (4,5) number and exposure time (s) of the VHR-V spectra; (6) red grism used; 
  (7,8) number and exposure time (s) of the red spectra.
\end{table}

\section{Observations and data analysis}
\label{observations}

We obtained new spectroscopic observations for five of the candidate relics,
while 3C~319 and 3C~052 could not be observed. With respect to the data
presented in \citet{buttiglione09}, we used longer exposure times and, in the
\Hb\ region, higher spectral resolution (corresponding to $\sim$400 \kms,
better matched with the intrinsic line widths) to improve the measurements of
the weak emission-lines characteristic of these objects. We also obtained new
red spectra for the three ELEGs in order to measure the intensity of the
[S~II] lines.

The spectra were taken with the Telescopio Nazionale Galileo (TNG), a 3.58 m
telescope located on the Roque de los Muchachos in La Palma Canary Island
(Spain). The observations were made with the DOLORES (Device Optimized for the
LOw RESolution) spectrograph. The detector used is a 2100x2100 pixel
back-illuminated E2V4240, with a pixel size of 0\farcs252. The observations
were carried out in service mode between February and September 2009. The
chosen long-slit width is 2$\arcsec$ and was aligned along the radio-jet axis
in order to increase the chance of detecting off-nuclear line emission, which,
if present, is more likely to be extended in this direction
\citep{baum89b}.\footnote{For the 3CR spectroscopic survey, the slit was instead
  aligned with the parallactic angle in order to minimize light losses due to
  atmospheric dispersion.}

For each target we generally took spectra with two grisms, VHR-V (4700-6700
\AA) and the VHR-R (6200-7700 \AA) or VHR-I (7400-8800 \AA), depending on
redshift, with a resolution of $R\sim$750, 1250, and 1500 respectively. In the
case of 3C~035, we instead used the lower resolution LRB grism ($\sim$3500-7700
\AA, R $\sim$250) but with an exposure time three times longer than that used
during the survey. The selected grisms cover the \Hb\ and \Ha\ spectral
regions in the observed frame. They include the most luminous emission-lines of
the optical spectrum and, in particular, the key diagnostic lines H$\beta$,
[O~III]$\lambda\lambda$4959,5007, [O~I]$\lambda\lambda$6300,64, H$\alpha$,
[N~II]$\lambda\lambda$6548,84, [S~II]$\lambda\lambda$6716,31. Exposures were
divided into at least two sub-exposures, obtained moving the target along the
slit. Table \ref{logoss} provides the journal of observations.

The data analysis was performed as described in \citet{buttiglione09} where
further details can be found. Summarizing, the spectra were bias subtracted
and flat fielded.  The spectra obtained in each sub-exposure were subtracted
to remove the sky background. The residual background was subtracted measuring
the average on each pixel along the dispersion direction in spatial regions
immediately surrounding the source spectrum. The data were then wavelength
calibrated and corrected for optical distortions. Finally, the spectra were
extracted and summed over a region of 2$\arcsec$ along the spatial direction
and flux calibrated using spectrophotometric standard stars, observed
immediately after each target.

\begin{landscape}
\begin{figure}[htbp]

\centerline{
\psfig{figure=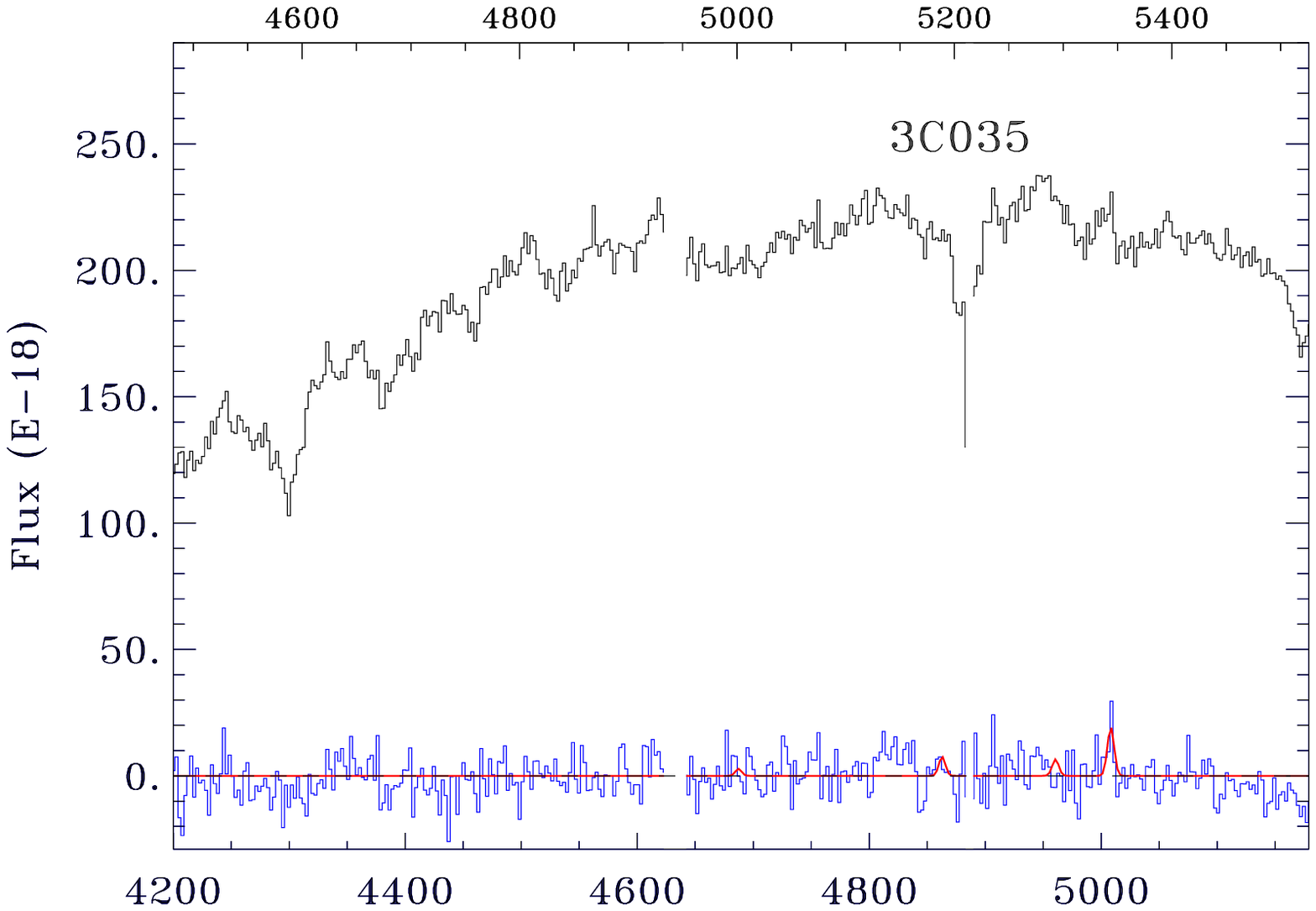,width=0.25\linewidth}
\psfig{figure=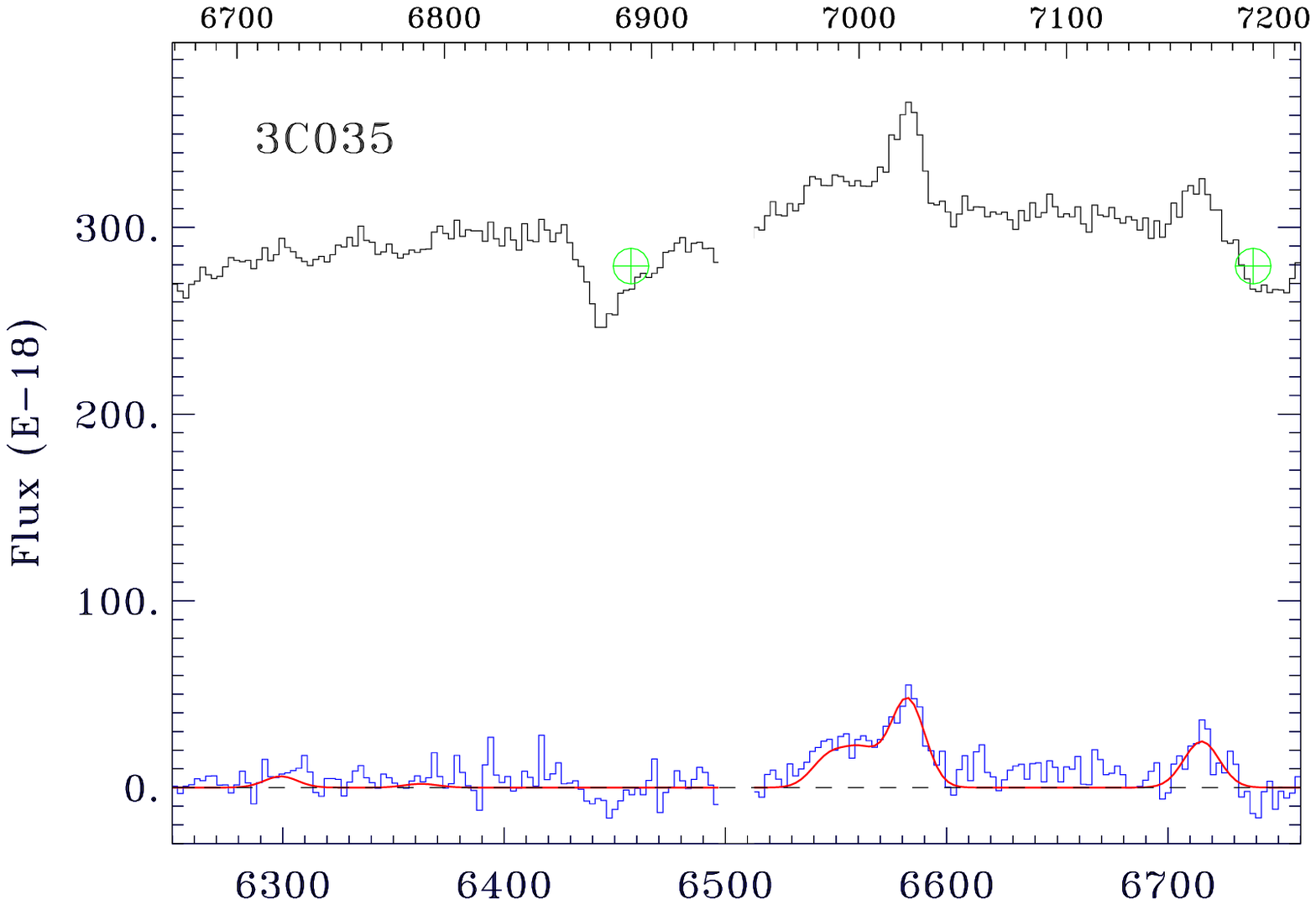,width=0.25\linewidth}
\psfig{figure=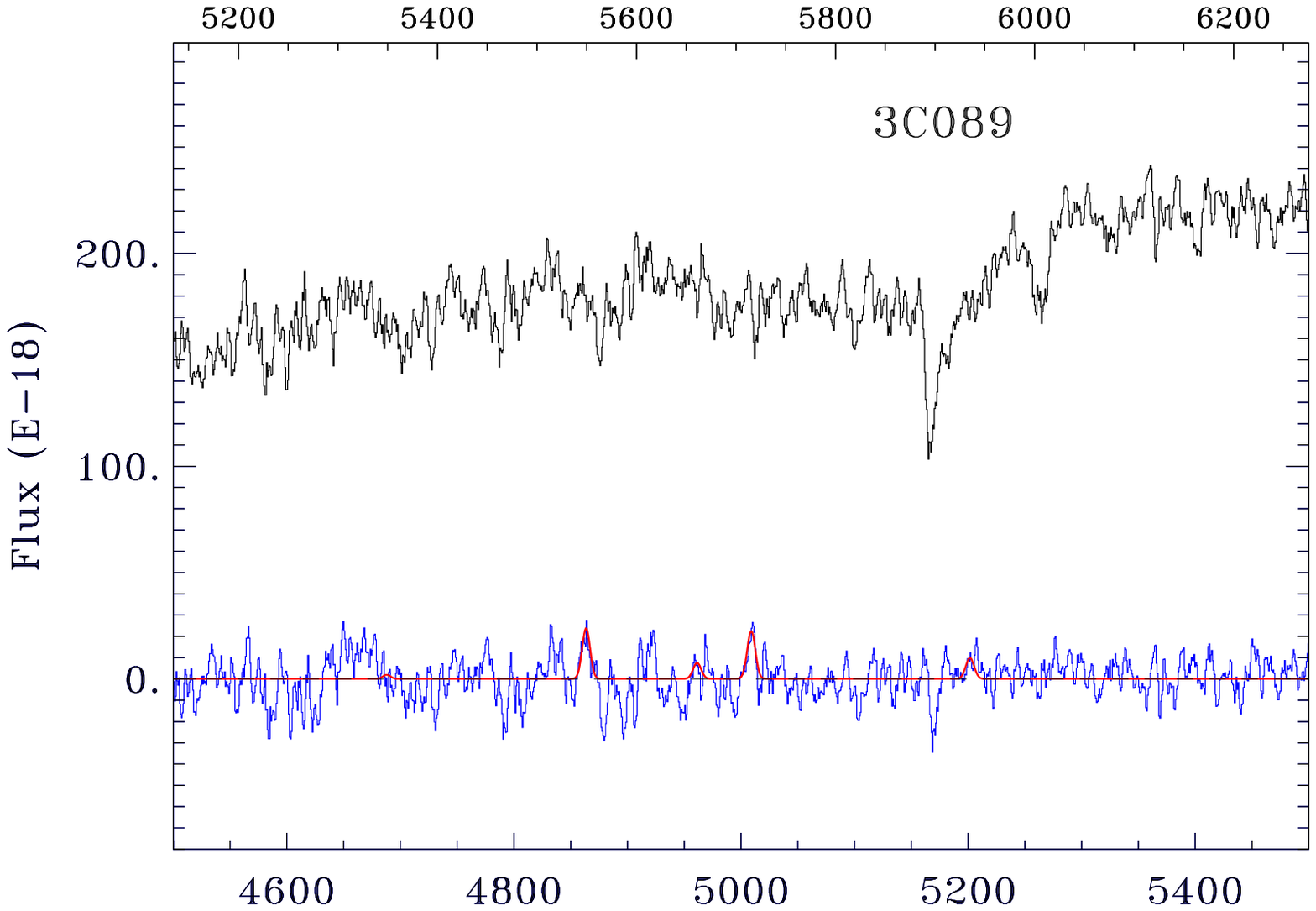,width=0.25\linewidth}
\psfig{figure=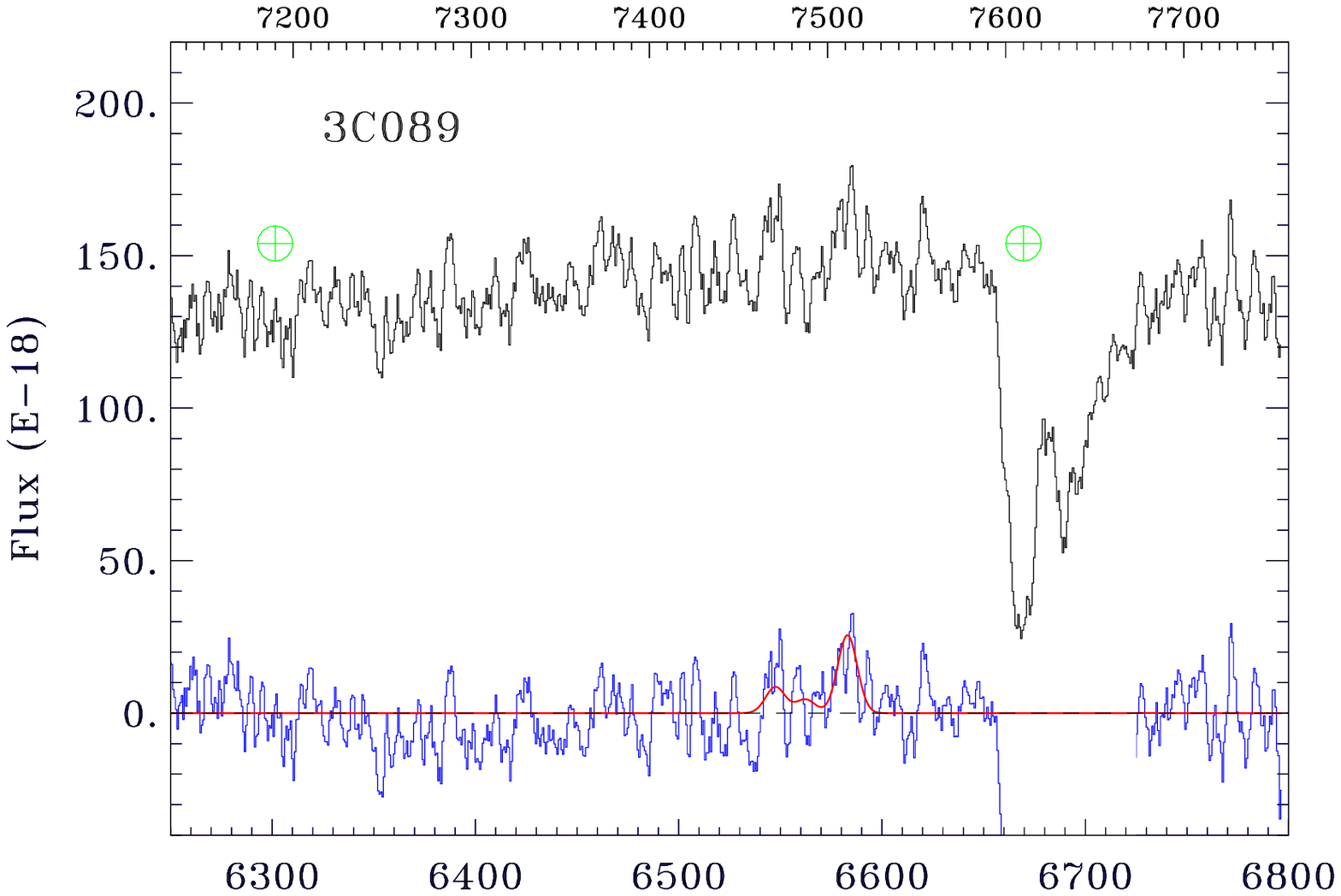,width=0.25\linewidth}}
\centerline{
\psfig{figure=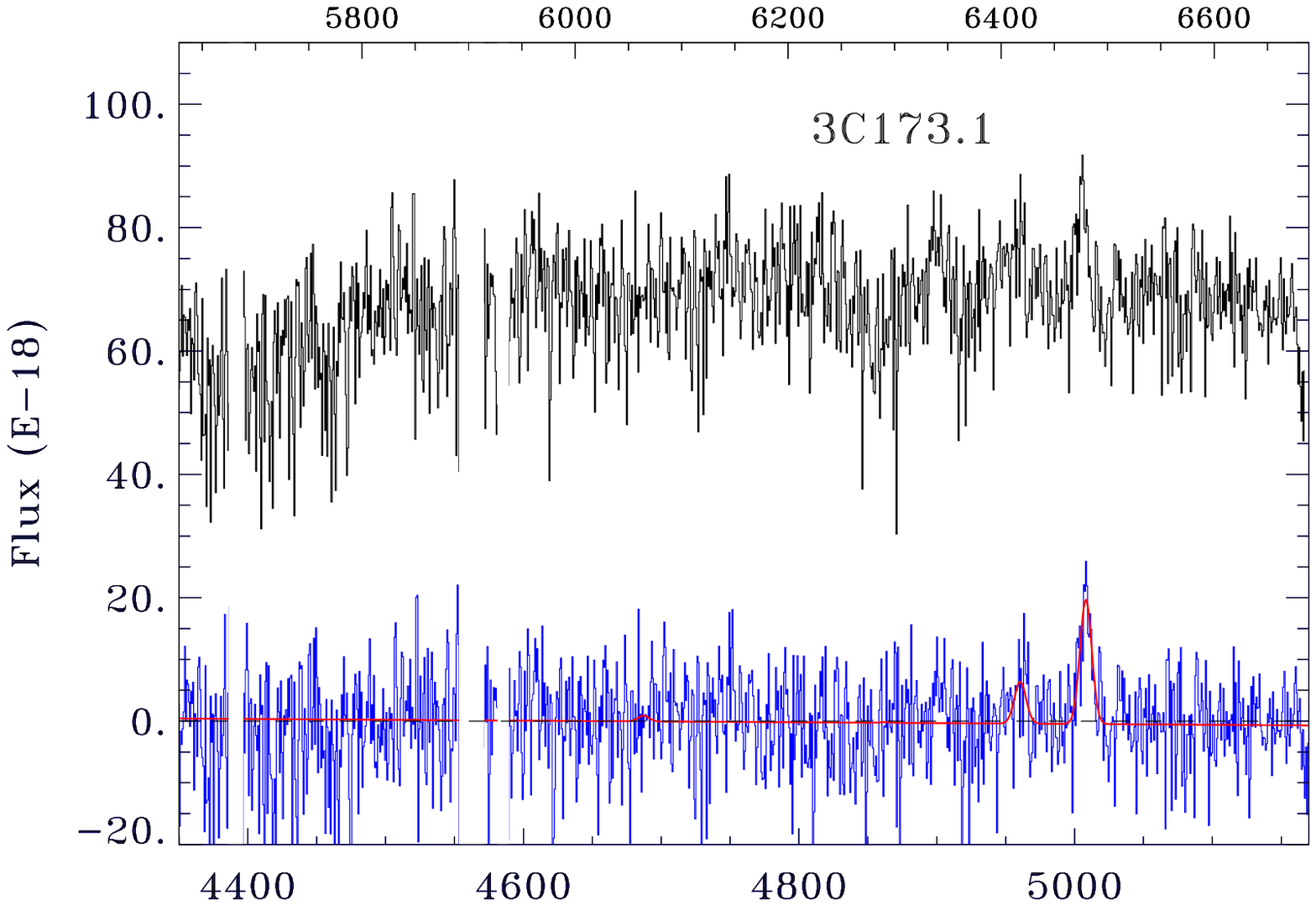,width=0.25\linewidth}
\psfig{figure=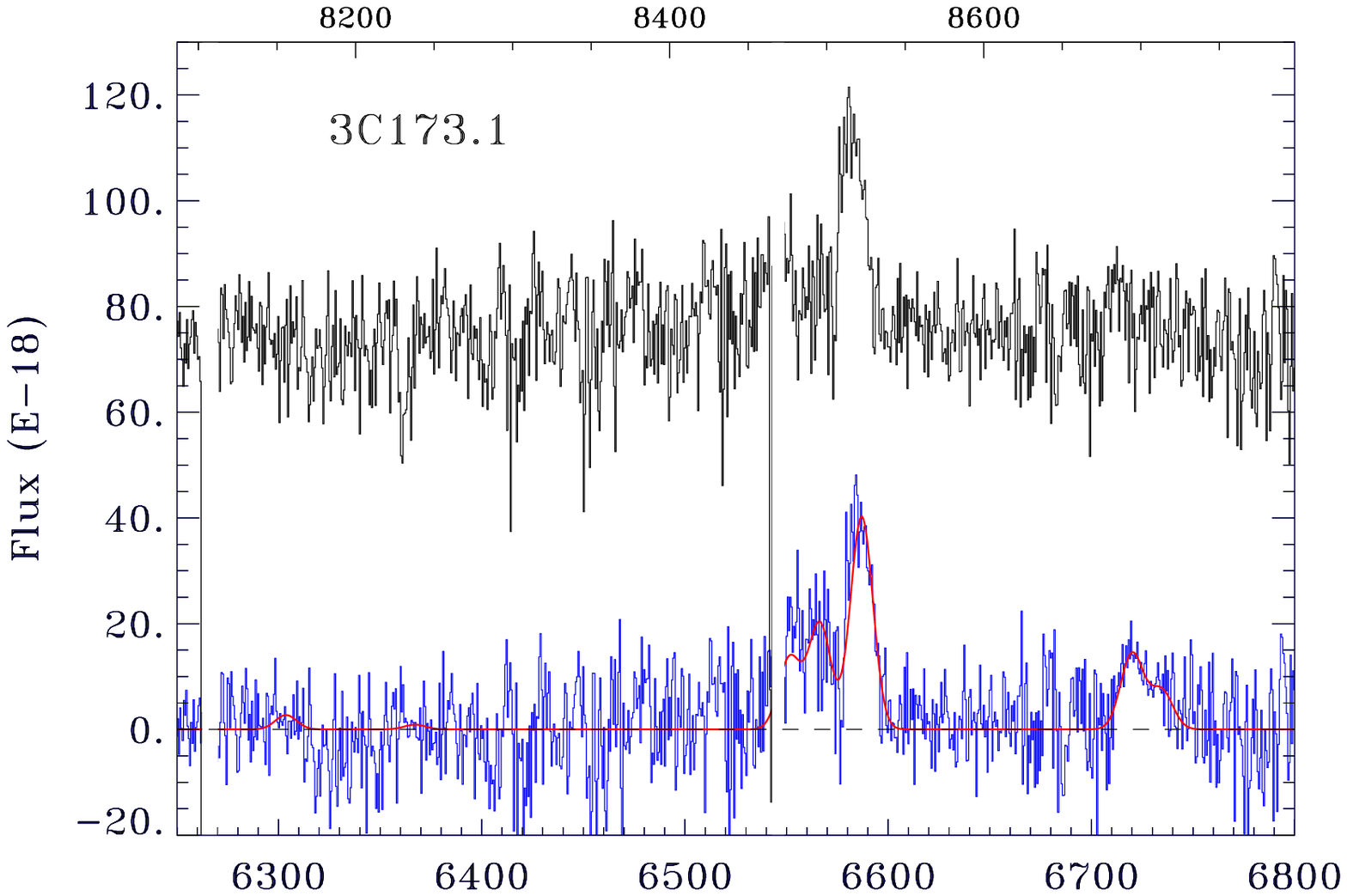,width=0.25\linewidth}
\psfig{figure=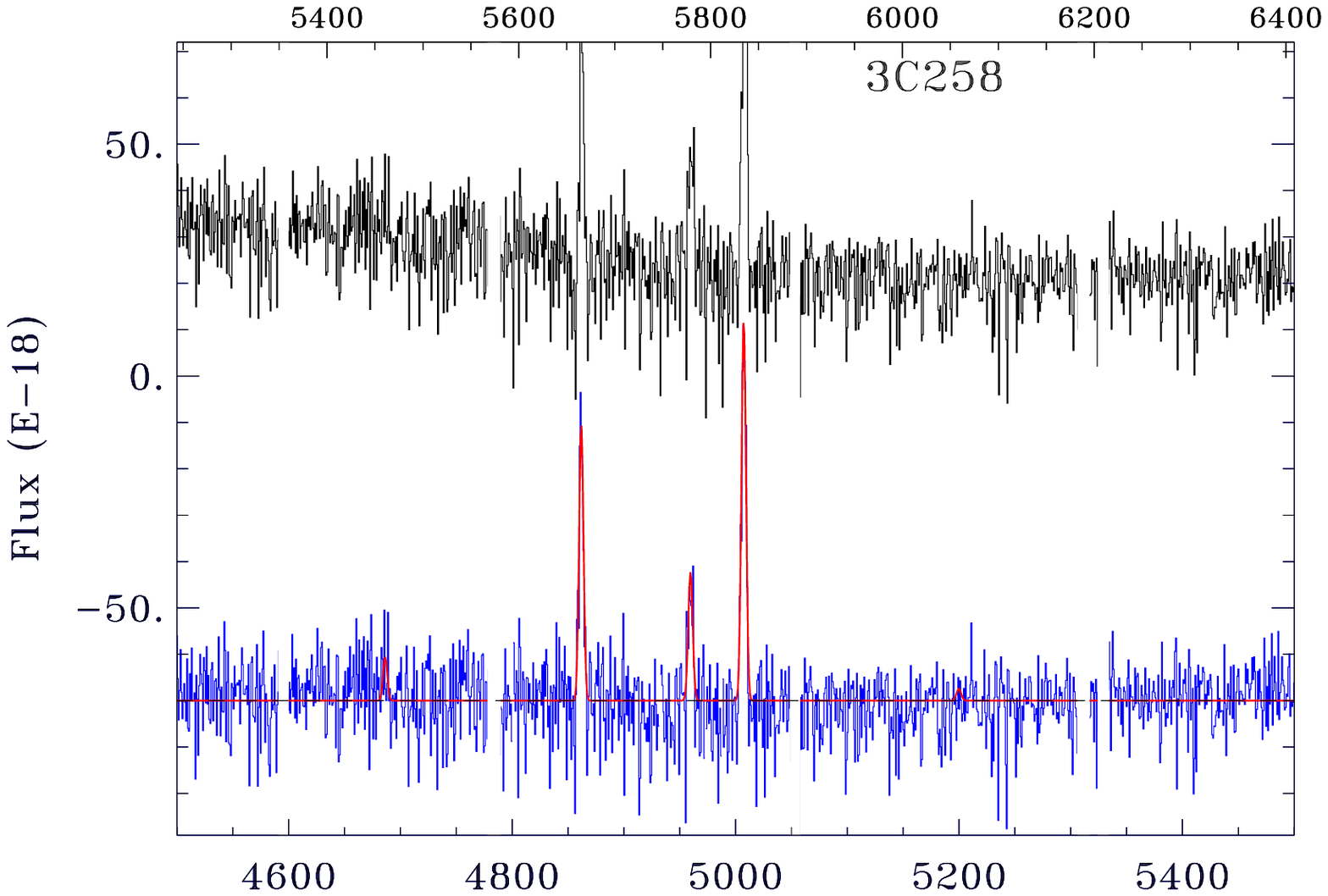,width=0.25\linewidth}
\psfig{figure=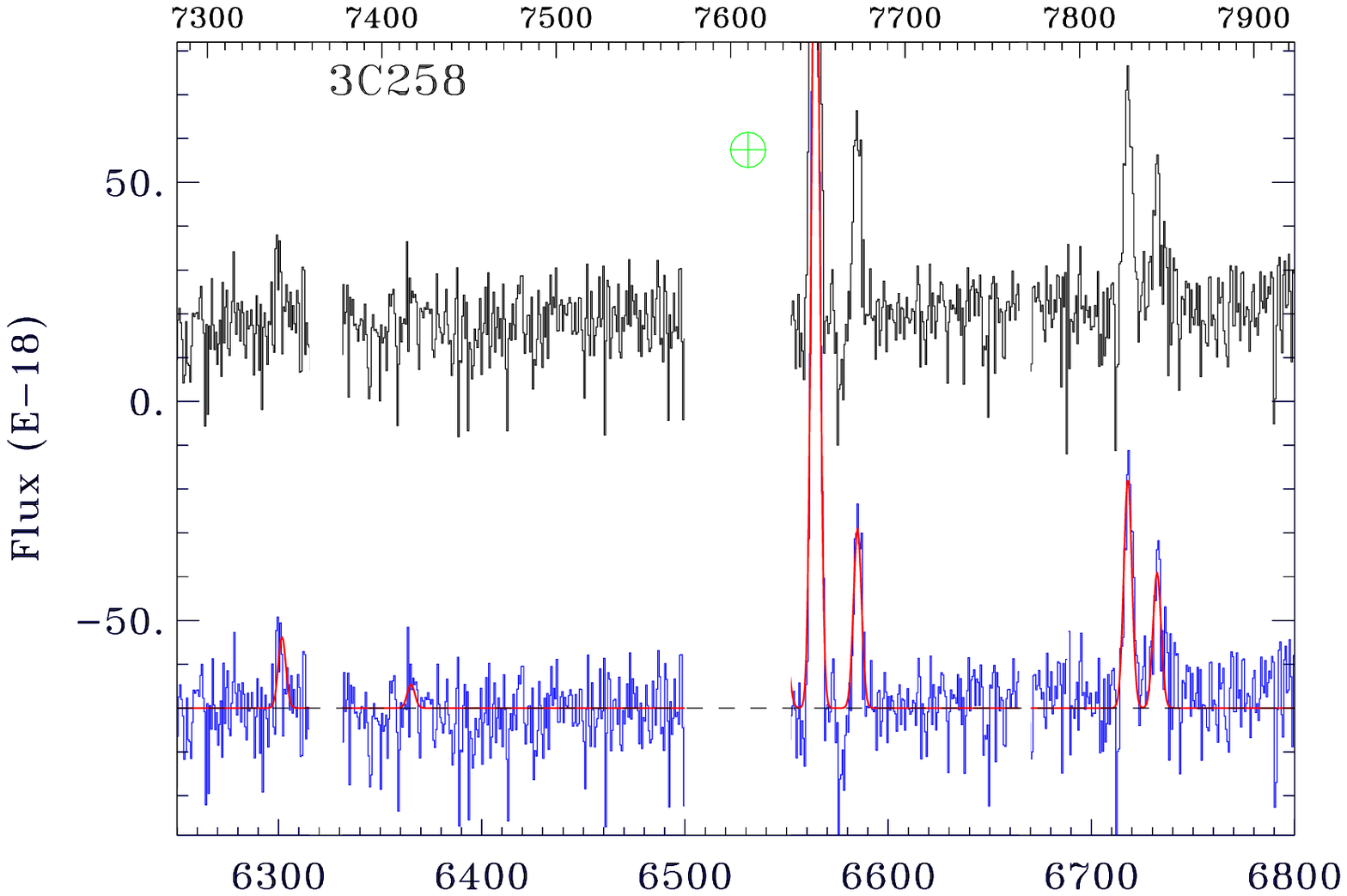,width=0.25\linewidth}}
\centerline{
\psfig{figure=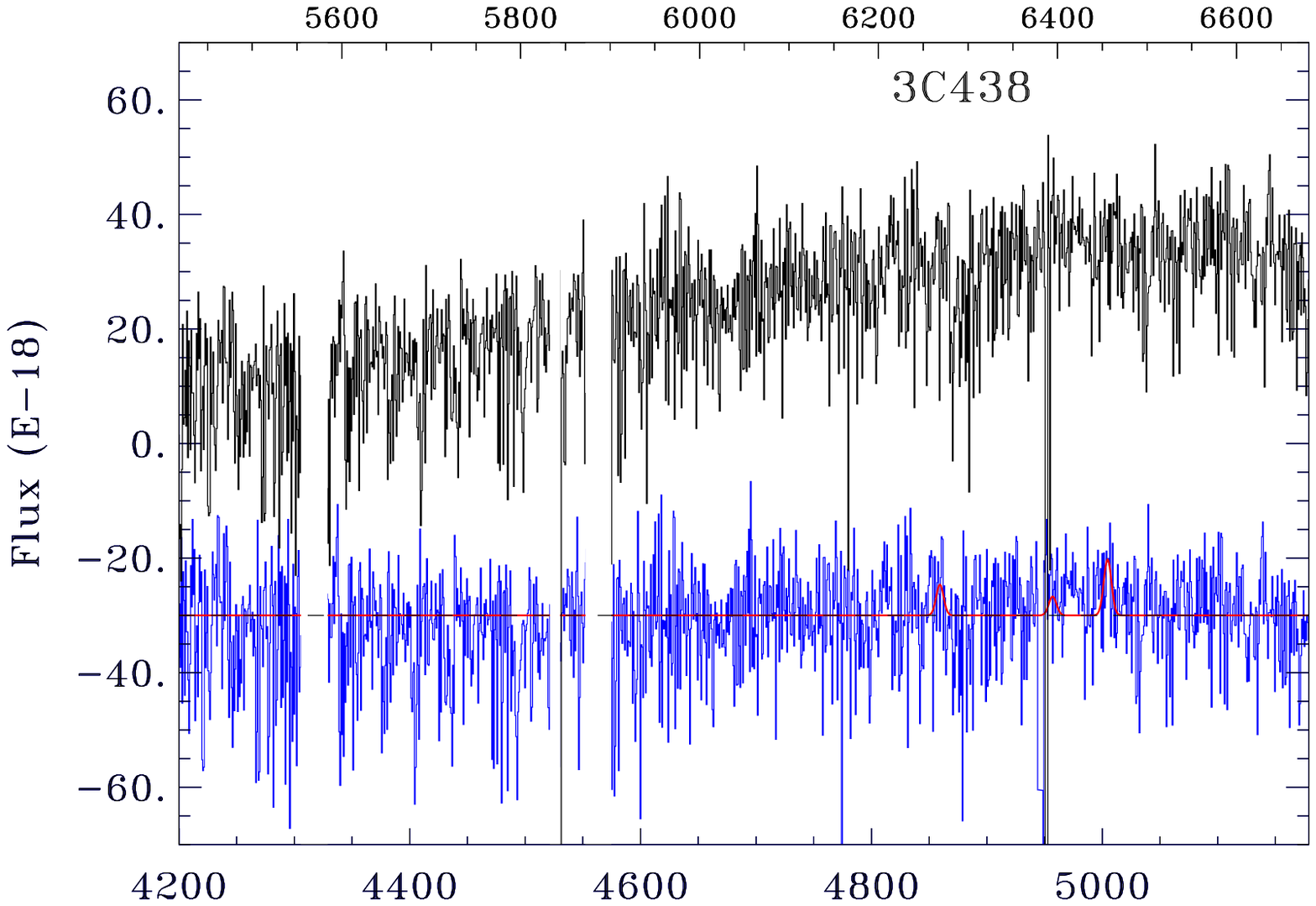,width=0.25\linewidth}
\psfig{figure=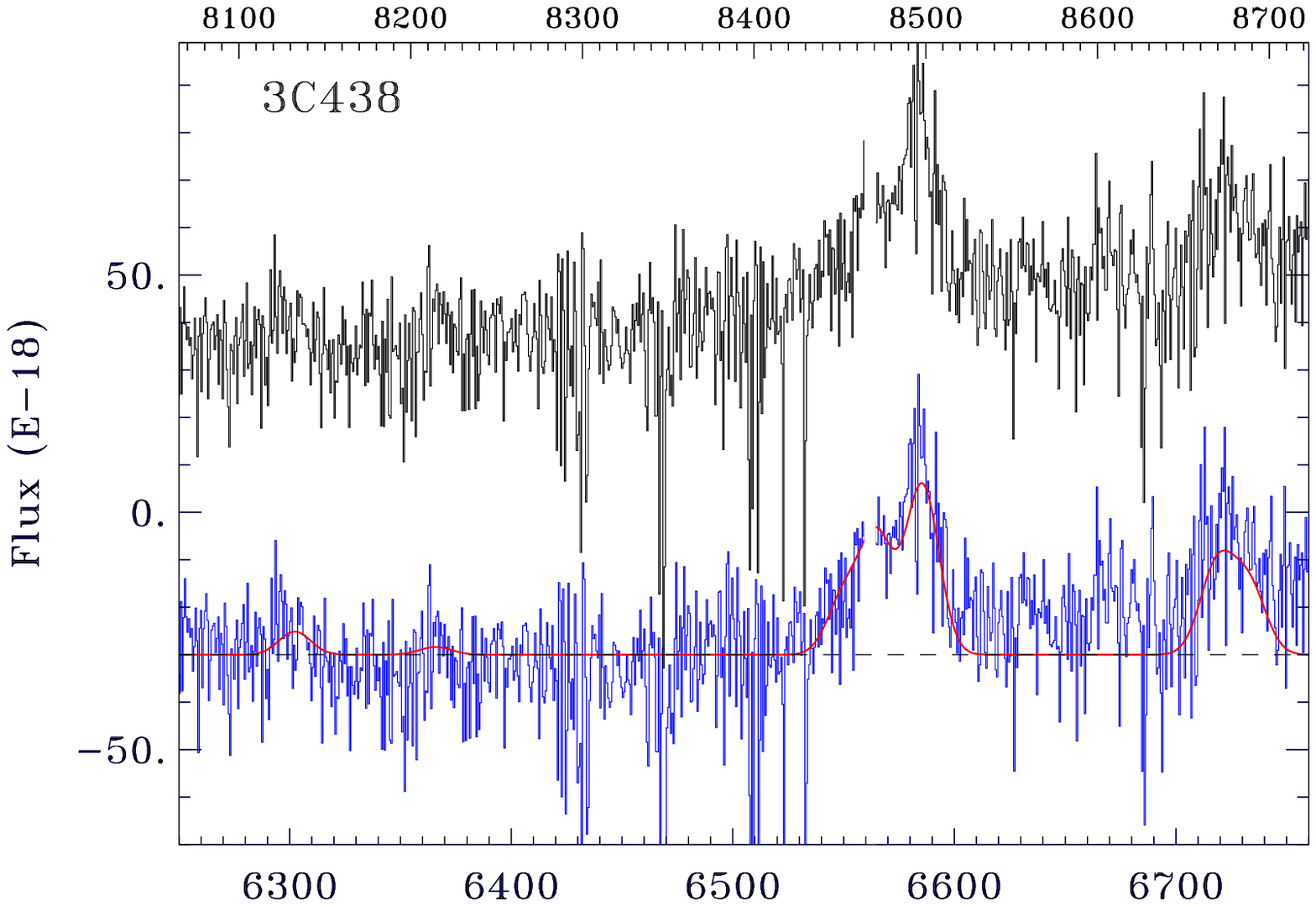,width=0.25\linewidth}}
\caption{\label{spectra2} For each of five the candidate relic RGs, the
  blue and red spectra are shown. The fluxes are in units of erg cm$^{-2}$
  s$^{-1}$ \AA$^{-1}$, while the wavelengths are in \AA. The upper axes of the
  spectra show the observed wavelengths while the lower axes show the rest
  frame wavelengths. The bottom blue spectra in each panel are the residuals
  after the subtraction of the stellar continuum; shown in red is the fit to
  the residuals which includes the emission-lines. For 3C~089, we rebin the
  spectrum to a 3\AA\ resolution to show more clearly the weak emission-lines. The main telluric absorption bands are indicated with circled
  crosses.}
\end{figure}
\end{landscape}

\begin{figure}[htbp]
\psfig{figure=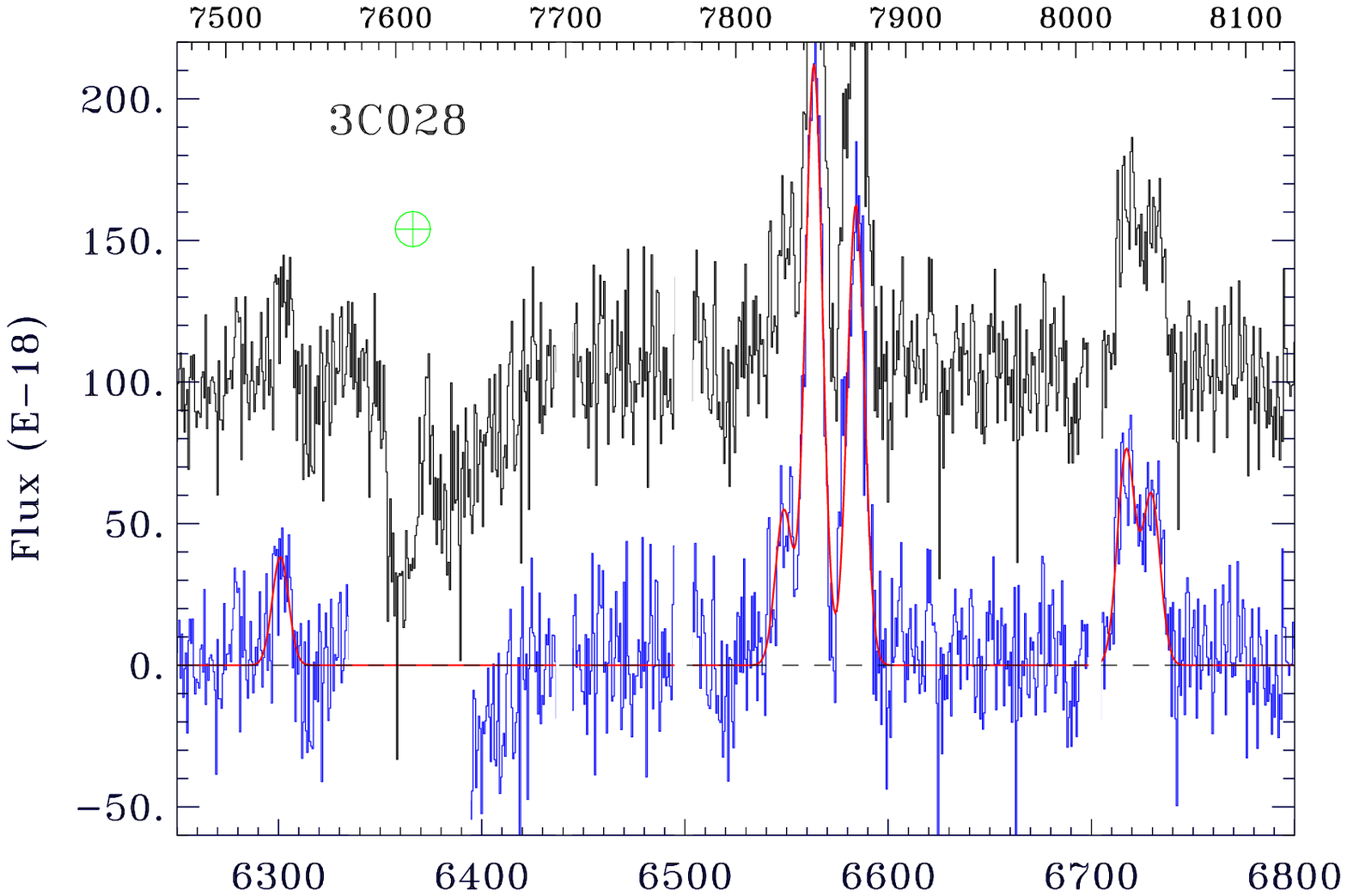,width=1.0\linewidth}
\psfig{figure=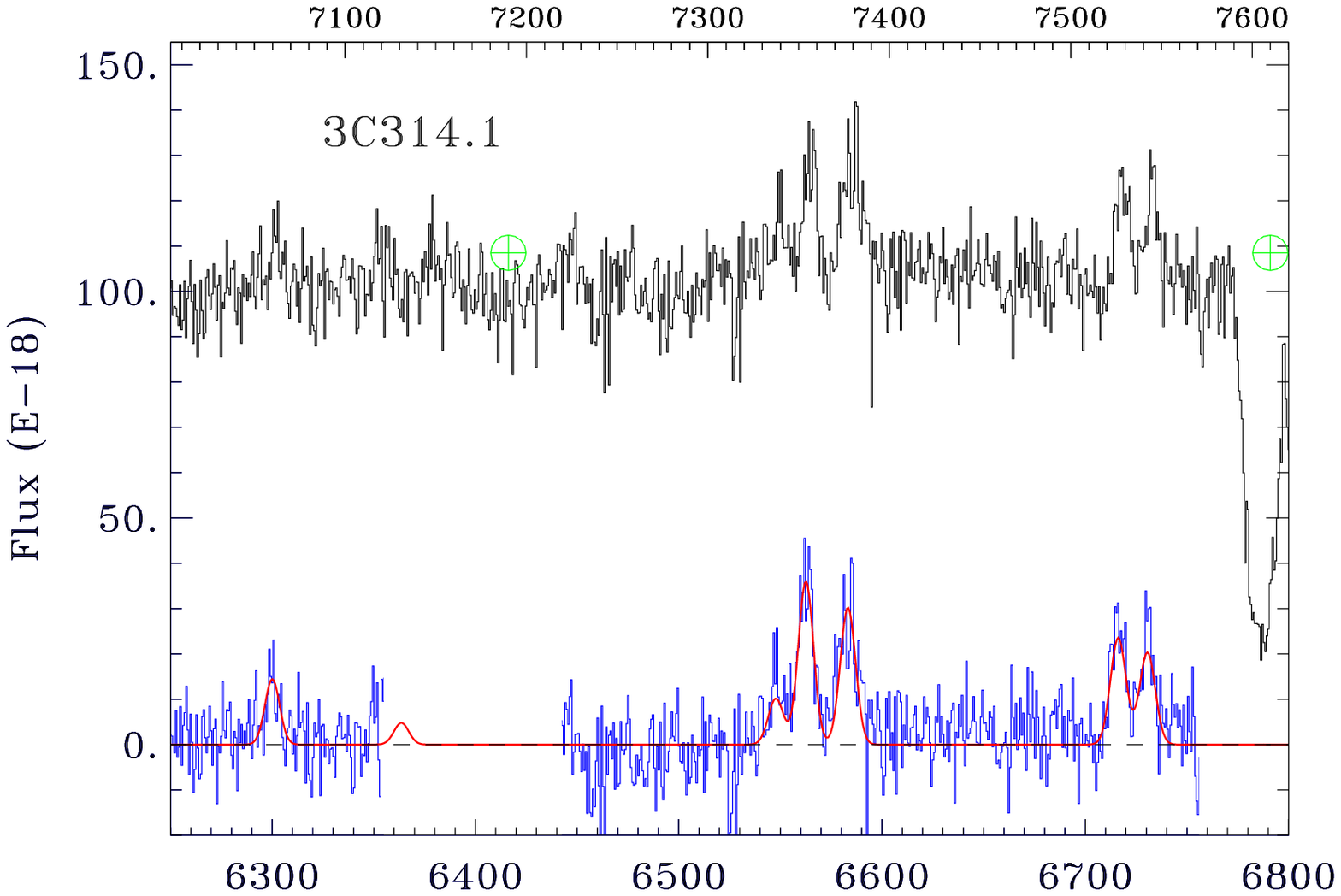,width=1.0\linewidth}
\psfig{figure=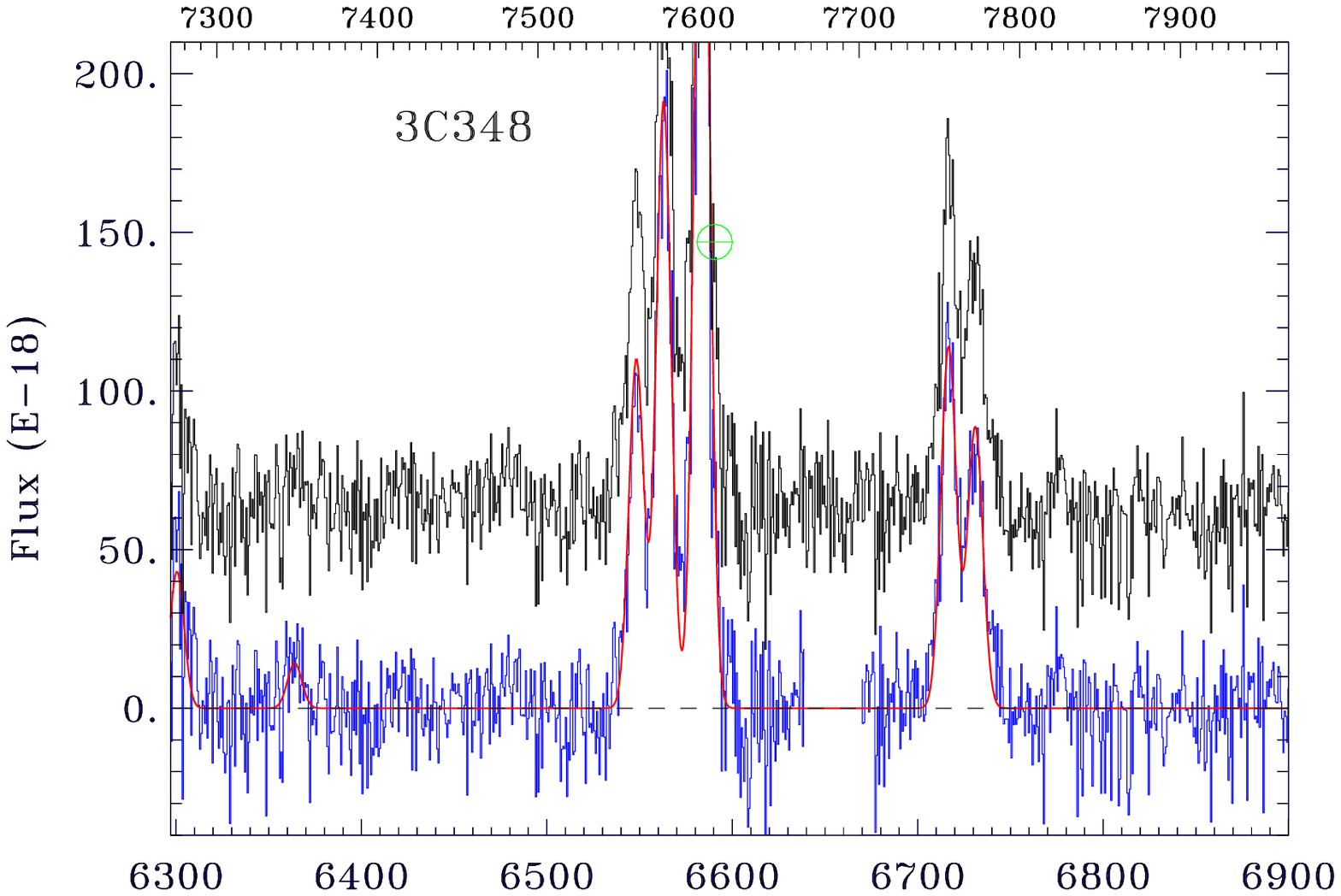,width=1.0\linewidth}
\caption{\label{spectra} New red spectra for the confirmed relics. The fluxes
  are in units of erg cm$^{-2}$ s$^{-1}$ \AA$^{-1}$, while the wavelengths are
  in \AA. The upper axes of the spectra show the observed wavelengths while
  the lower axes show the rest frame wavelengths. The bottom spectra in each
  panel are the residuals, after the subtraction of the stellar continuum;
  shown in red is the fit to the residuals which includes the emission-lines.
  The main telluric absorption bands are indicated with circled crosses.}
\end{figure}

For all the observed targets Figures \ref{spectra2} and \ref{spectra} show the
resulting calibrated spectra, corrected for reddening due to the Galaxy
\citep{burstein82,burstein84} taken from the NASA Extragalactic Database (NED)
and using the extinction law of \citet{cardelli89}, and transformed
into rest frame wavelengths.

The contribution of stars to the nuclear spectra was subtracted using the
best-fit single stellar population model taken from the \citet{bruzual03}
library. We excluded from the fit the spectral regions corresponding to
emission-lines, along with other regions affected by telluric absorption,
cosmic rays, or other impurities. By using the

\noindent {\it specfit} package in IRAF, we then measured the line intensities
adopting Gaussian profiles.  We required the widths and the velocity to be the
same for all the lines within each spectrum. The lines fluxes were free to
vary, except for those with known fixed ratios.

Table \ref{bigtable} summarizes the intensities and the errors of the main
emission-lines (de-reddened for Galactic absorption) relative to the intensity
of the H$\alpha$. We placed upper limits at a 3$\sigma$ level to the
undetected, but diagnostically important, emission-lines by measuring the
noise level in the regions surrounding the expected positions of the lines
and adopting as line width the instrumental resolution.

\begin{table*}
\begin{center}
\caption[Emission-line measurements.]{Emission-line measurements.} 
\label{bigtable} 
\begin{tabular}{l | c c c c c c }
\hline\hline
Name     & H$\beta$ & [O III]$\lambda$5007 & [O I]$\lambda$6300 & [N II]$\lambda$6584 & [S II]$\lambda$6716 & [S II]$\lambda$6731 \\
\hline	
3C~028   & \multicolumn{2}{c}{}  & 0.17 (26) & 0.77 (10) & 0.36( 6) & 0.29( 6) \\
3C~314.1 & \multicolumn{2}{c}{}  & 0.38 (14) & 0.84 ( 6) & 0.67( 8) & 0.58( 9) \\
3C~348   & \multicolumn{2}{c}{}  & 0.22 ( 9) & 1.72 ( 1) & 0.61( 3) & 0.47( 6) \\
\hline                                
3C~035   &$<$0.47 &0.39 (26)& $<$0.82& 2.60 ( 8) &1.38(14) &--- \\
3C~089   &0.27 (31) &0.28 (26)& $<$2.10& 5.72 (13) & --- &--- \\
3C~173.1 &$<$0.25 &0.83 (11)& $<$0.47& 2.02 ( 7) &0.73(17) & 0.38 (29) \\
3C~258   &0.30( 6) &0.42 ( 4)& 0.08 (20)& 0.20 ( 8) &0.26( 6) & 0.16 (10) \\
3C~438   &$<$0.38 &$<$0.54& 0.19 (23)& 1.46 ( 6) &0.78 ( 8) & 0.60 ( 8) \\
\hline
    \end{tabular}                                               
  \end{center}                                                  
  Column description: (1) source name; (2 through 7) 
  de-reddened flux ratios with respect to H$\alpha$.  
  The values in parentheses report the errors (in percentage) of each line.
\end{table*}

\section{Results}
\label{sect2}

\subsection{Comments on the individual relic candidates}

\begin{figure*}[tp]
  \centerline{ \psfig{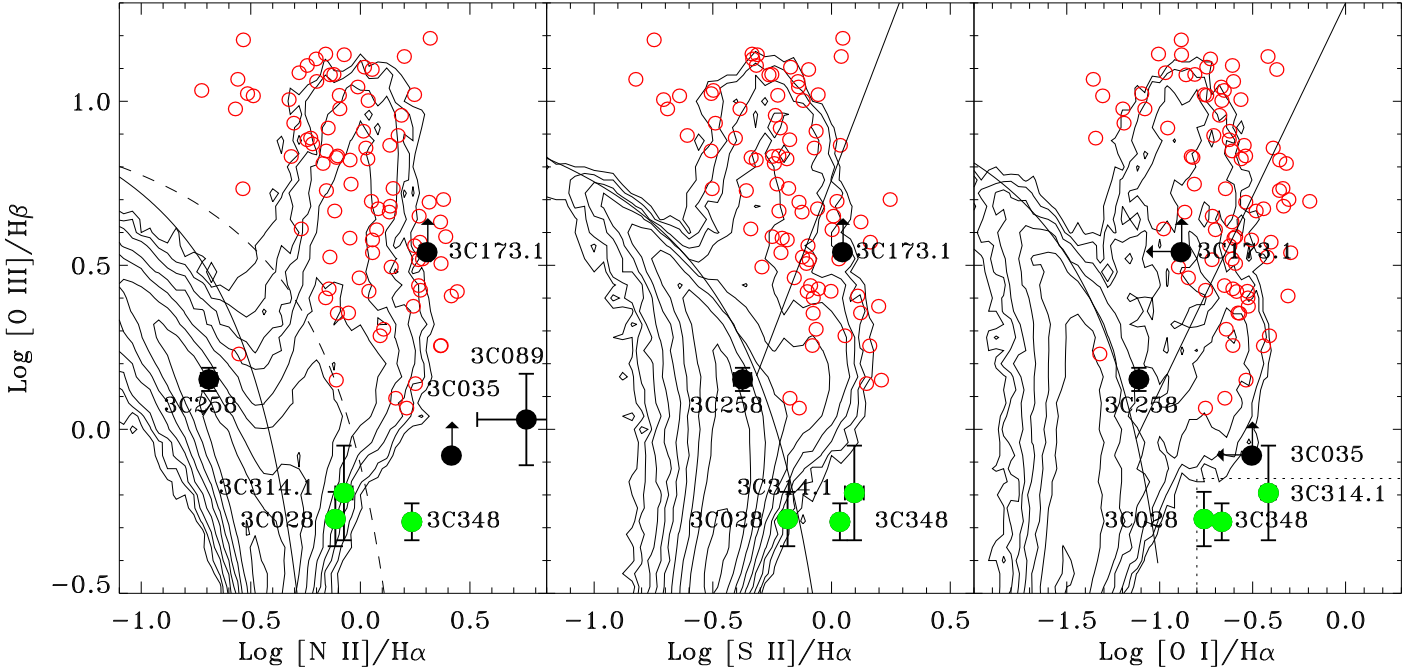}}
  \caption{\label{diag} Contours: density of SDSS emission-line galaxies in
    the optical diagnostic diagrams (adapted from \citealt{kewley06}); only
    galaxies with lines measured with a signal-to-noise ratio $\geqslant$ 5
    are included.  Levels are in geometric sequence with a common ratio of 2.
    Galaxies below the curved solid lines are star-forming galaxies. In the
    left diagram, the dashed curve marks the transition from composite galaxies
    to AGN; in the middle and right panels, the oblique line separate LINER
    from Seyfert. The empty circles are the 3CR sources with z$<$0.3, while
    the filled circles are the ELEGs (green) and the candidates (black). In
    the right diagram, the region bounded by the dotted lines defines the
    ``relic neighbors'' (see Sect. \ref{lifetime}).}  \end{figure*}

\noindent {\bf 3C~035}: The \Hb\ line remains undetected in the new spectrum.
With respect to \citet{buttiglione10} the lower limit to the [O~III]/\Hb\
ratio increases to log [O III]/\Hb\ $> -0.08$ (from a previous limit of
$>-0.30$), slightly higher than that measured in the three ELEGs, but still lower
than any other 3CR source.

\noindent {\bf 3C~089}: While from the survey data we were only able to detect
the [N~II] and \Ha\ lines, we can now also measure the fluxes of \Hb\ and
[O~III], with a ratio of log [O III]/\Hb\ = 0.03. This places 3C~089 in the
first spectroscopic diagram (left-hand panel in Fig. \ref{diag}) toward the
bottom of the locus populated by the 3CR RGs, and well separated from the
ELEGs.

\noindent {\bf 3C~173.1}: The upper limit to its \Hb\ flux corresponds to a
lower limit of log [O~III]/\Hb\ $>$ 0.54, sufficient to 
establish that this source is not an ELEG.

\noindent {\bf 3C~258}: The line ratios are characteristic of a star-forming
galaxy. Indeed, in two out of three diagnostic diagrams (see Fig. \ref{diag}),
3C~258 is located below the solid lines separating AGN from star forming
galaxies while in the third diagram, ([O~I]/\Ha\ vs. [O~III]/\Hb) is only
marginally above the boundary. This is the second 3CR source (at z $<$ 0.3),
with 3C~198, that shows these spectroscopic properties. 3C~198 is a large
classical double, while 3C~258 is a compact steep spectrum (CSS) source with a
double morphology, barely resolved in MERLIN maps \citep{akujor91}. The host
morphology is peculiar \citep{floyd08} because it is very compact for its
redshift\footnote{There is some controversy on the redshift of 3C~258. Our
  data confirm the value of 0.165 of \citet{smith76}.}  and has an arc-like
structure toward SE that is possibly the sign of a recent merger.

\noindent {\bf 3C~438}: We still cannot detect the \oiii\ and the \Hb\
lines. No classification is possible for this source.

\medskip

\begin{figure*}[htbp]
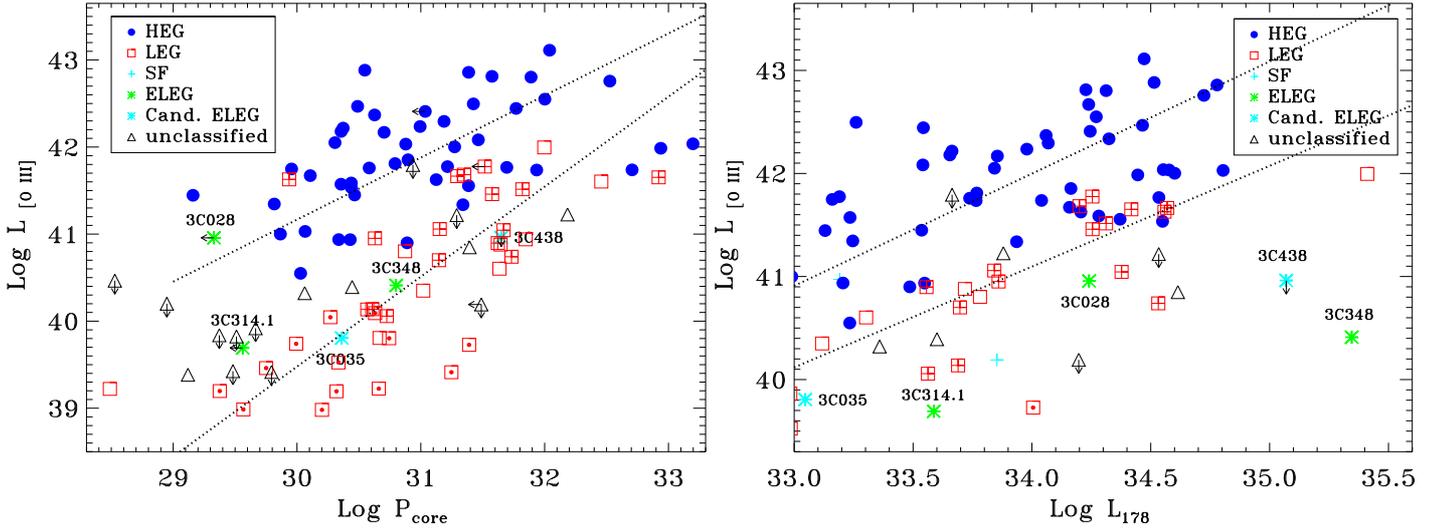
 
\centerline{
\psfig{figure=20662f4a.epsi,angle=90,width=0.5\linewidth}
\psfig{figure=20662f4b.epsi,angle=90,width=0.5\linewidth}}
\caption{\label{lo3pcore} [O III] luminosity (\ergs) versus (left) core power
  and (right) total 178 MHz radio power (\ergsHz) for the 3CR sources with
  z$<$0.3 limiting, for more clarity, to those with Log L$_{178} > 33$. We
  mark LEGs with FR~I (FR~II) morphology with a small circle (cross) within the
  large square.}
\end{figure*}

In Fig. \ref{diag} we show the locations of the three ELEGs and candidate
relics in the emission-lines diagnostic diagrams, compared to the rest of the
3CR sample with z$<$0.3 and superimposed on the contours representing the
density of the SDSS emission-line sources \citep{kewley06}. The newly measured
line ratios for three of the candidate relics locate them outside the region
characteristics of ELEG (3C~089 is a LEG, 3C~173.1 could be either a LEG or a
HEG, and 3C~258 is star-forming galaxy). They are insufficient to establish
the nature of the remaining two, 3C~035 and 3C~438, which we still consider as
candidate ELEGs. Nonetheless, they are located in the same region of the
[O~III] luminosity - radio core power diagram (Fig. \ref{lo3pcore}, left-hand
panel) as the LEGs. Their core dominance is also similar to the values
reported above for the 3CR sources (F$_{\rm core}$/F$_{\rm tot} \sim10^{-2.7}$
and $\sim10^{-3.4}$ for 3C~035 and 3C438, respectively). However, they both
show a deficit in line emission with respect to the total radio luminosity,
which is particularly strong (by a factor of $\gtrsim$10) in the case of 3C~438.

\subsection{Gas density measurements in the ELEGs}

The new result for the three galaxies already classified as ELEG is an accurate
measurement of the ratio between the two lines forming the [S~II] doublet,
from which it is now possible to measure the gas density. Following
\citet{osterbrock89}, we estimated the gas electron density by using the ratio
$R_{\rm [S~II]}=$ [S~II]$\lambda$6716/[S~II]$\lambda$6731. The resulting
values are $n_e = 190^{+110}_{-80}$ cm$^{-3}$ for 3C~028, $300^{+250}_{-120}$
cm$^{-3}$ for 3C~314.1, and $160^{+90}_{-70}$ cm$^{-3}$ for 3C~348.  We
assumed a temperature of 10$^4$ K, but the derived density depends only weakly
on temperature, as $T^{-1/2}$.

As a comparison we consider the gas density in the rest of the 3CR sample.
There are 73 sources in \citet{buttiglione09,buttiglione11} where the [S II]
line ratio can be measured with an accuracy better than 10\%. The median
$R_{\rm [S II]}$ value is 1.15, corresponding to $n_e\sim300$ cm$^{-3}$, and
60\% of them have $1.0 < R_{\rm [S II]} < 1.35$ ($100 \lesssim n_e \lesssim
600$ cm$^{-3}$). Thus the gas density in relics is not strikingly different
from the rest of the sample.

\section{Time evolution of line luminosities and ratios}
\label{sect3}

After measuring the gas density for the three relics, we used the analysis of
the time evolution of a photoionized cloud following an instantaneous drop of
the ionizing photon field described by \citet{binette87}. The decay time (the
time after which the line intensity decreases by a factor 1/$e$) of the
[O~III] line can now be estimated, resulting in $t_d({\rm [O~III]}) \sim20 \,
n_{e,2}^{-1}$ years, where $n_{e,2}$ is the gas electron density in units of
$10^2$ cm$^{-3}$.  The new density measurements set this parameter to $n_{e,2}
\sim$ 1.6 - 3 for the three ELEGs.  The decay time of the \Hb\ line is instead
$t_d ({\rm H}\beta) \sim1300 \, n_{e,2}^{-1}$ years.

The actual NLR structure is more complex than the single cloud of constant
density used in the models by \citeauthor{binette87}, because it is formed by
clouds of different densities distributed at different distances over the
central regions of the galaxy. The information about the decrease in the
ionizing luminosity of the AGN reaches these clouds at different times
depending on their distance from the nucleus.  As long as the time elapsed
since the AGN ``switched off'' is less than the NLR light-crossing time, the
NLR is divided into two regions, the outer one with a spectrum still
characteristic of a galaxy in an active state and an inner portion with a
``fossil'' spectrum. The thickness of the intermediate layer, given by
$t_d{\rm [O~III]} \times c$, is only 2 - 4 pc, much smaller than both the NLR
size and the region covered by the slit used for our observations,
which is 2 - 3 kpc.

\begin{figure*}[htbp]
\centerline{
\psfig{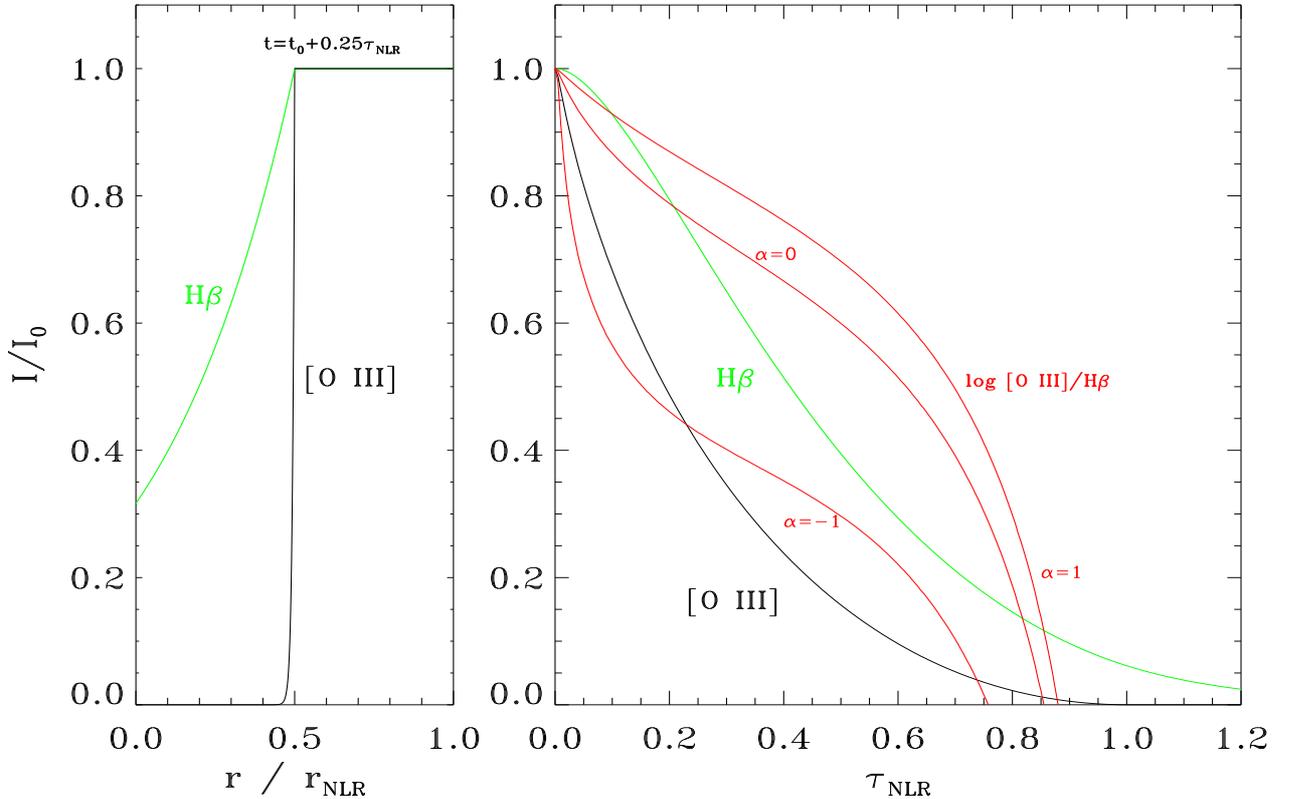}}
\caption{\label{evol} Temporal evolution of the [O~III] and \Hb\ intensity
  following a drop in the nuclear radiation field for a spherical NLR of
  radius 1 kpc, with an initial emissivity law dropping as $r^{-2}$, and a
  density $n_e=100 \,{\rm cm}^{-3}$. Left panel: emissivity at each radius
  after $\sim$1500 years. Right panel: temporal evolution of the total lines
  intensity and of their ratio. We also report the changes of ratio with time
  for different emissivity laws in the form of
  $r^{-2+\alpha}$, with $\alpha = \pm 1$.}
\end{figure*}

This implies that the NLR temporal evolution is dominated by geometric
effects. Indeed the observed spectrum is the superposition of the ``active''
and the ``fossil'' regions, while the contribution of the intermediate layer
is negligible. The observed line ratios are then determined by the relative
amount of line emission in the two regions. This depends on the distribution
of ionized gas within the NLR. A simple toy model, in which the line
emissivity drops as $r^{-2}$ and extends to a radius $r_{\rm NLR} = 1$ kpc,
can be used to follow the NLR evolution. In this case, the NLR can be thought
of as the sum of shells of equal thickness, each producing the same amount of
line emission.  The \oiii\ emissivity in each shell drops rapidly following
the change in the flux of ionizing photons and the total \oiii\ flux declines
(see Fig. \ref{evol}). The \Hb\ also decreases, but due to its longer decay
time, at a slower rate.  However, from the point of view of a distant observer,
each point (r, $\theta$) in the NLR only responds to the drop in ionizing flux
after a delay $\Delta t = r/c(1-cos\theta)$, where $\theta$ is the polar angle
measured from the line of sight. The boundary between the ``active'' and
``relic'' regions is therefore a parabolic isodelay surface with its symmetry
axis along the line of sight and its focus at the AGN (e.g.,
\citealt{bahcall72}).

The resulting \oiii/\Hb\ ratio, which we initially set to 10, also lowers and
is reduced to 0.5 (the value observed in the ELEGs) after a time interval
only slightly smaller than the NLR light travel time, $\tau_{\rm NLR} = 2
\times r_{\rm NLR}/c$. At this epoch, the \Hb\ intensity is reduced by a
factor of $\sim$10. These are rather general results and depend only slightly
on the adopted gas density or emissivity law with radius (see
Fig. \ref{evol}). The conclusion is that the typical timescale of the
evolution is the light travel time across the NLR and not the decay time of
the various lines.

For this analysis we assumed an instantaneous drop of the nuclear
source. However, as discussed by \citeauthor{binette87}, effectively the same
results are obtained if the decline occurs over less than $\tau_{\rm NLR}$
years.

\section{Relics versus active radiogalaxies}
\label{lifetime} 

Let us now assume that relic RGs are the result of a sudden and complete
switch-off of the nuclear activity. In this case the fraction of relics (with
respect to the galaxies in the active phase) is given by the ratio between the
duration of the transition phase and the lifetime of RGs, $\tau_{\rm RG}$. As
reported in \citet{relic} the fraction of relics can be estimated once all RGs
in a complete sample are properly spectroscopically classified, a situation
that is (almost) met in the 3CR sample with z$<$0.3. Considering the range of
radio luminosity where relic RGs are observed, Log L$_{178}$ = 33.5 -- 35.5
\ergsHz, there are 58 objects in the 3CR sample up to a redshift of 0.3: 35
HEGs, 16 LEGs, two star-forming galaxies, and three relics. The last two
objects are the remaining candidate relics (3C~035 and 3C~438), which are
still not classified after the analysis of new observations presented in
Sect. \ref{sect2}. The fraction of relics is then at least 3/58 ($\sim$5\%)
and can increase to $\sim$8\%, depending on the nature of the two unclassified
sources. A further uncertainty is related to the relic progenitors, i.e.,
depending on which spectroscopic class can evolve into a relic: HEG, LEG, or
both. However, this can only increase the relic fraction.

As discussed in the previous section, the duration of the transition phase is
a small fraction of the light travel time across the NLR (or, more precisely,
of the portion of the NLR covered by our spectra, 2 - 3 kpc), i.e., a few
thousand years. This line of reasoning would imply a lifetime of
$\sim 10{^5}$ years for RGs. This value is unacceptable from several points of view.
The most straightforward is the contrast with the median size of the FR~II RGs
in our sample, $d_{\rm RG}\sim$250 kpc, which exceeds the maximum size
attainable in such a timescale. A conservative upper limit to the fraction of
relics can be estimated as $c \times \tau_{\rm NLR} / d_{\rm RG} <
10^{-2}$. Thus, the transition phase is too short to correspond to a
substantial population of relic RGs.

Nonetheless, there are at least two alternative scenarios, always within the
framework in which the NLR evolution is governed purely by photoionization
from the nuclear source. The first one is that the phase of low nuclear
activity is transient. In this case  the relative number of active/relic galaxies should be
taken as an estimate of the ratio of the number of objects in
high/low states.

In the second one, the relic RGs are simply objects in which the NLR
ionization level is intrinsically very low, representing the tail of the broad
distribution of the NLR ionization parameter. Photoionization models indicate
that the observed \oiii/\Hb\ ratio can indeed be realized (see, e.g.,
\citealt{kewley06}).  However, as noted above, the relic RGs fall in regions
of the diagnostic diagrams scarcely populated by SDSS emission-line
galaxies. Considering the [O~I]/\Ha\ vs. \oiii/\Hb\ plane, the ``relic
neighbors'', i.e, the SDSS sources falling in the region bounded by the dotted
lines in Fig. \ref{diag} (right panel), are only 0.06\% of the SDSS
emission-line galaxies and 0.3\% of the $\sim$ 12,000 galaxies falling in the
AGN region. Within the AGN population, the line ratios observed in the relics
are very rarely achieved and, together with the peculiar radio properties,
appear instead to be the signposts of their distinctive evolutionary state.

\section{Origin of line emission in relics}
\label{origin} 

The results presented above argue against the scenario in which the population
of relic RGs correspond to objects that underwent a drop of nuclear source
intensity (or a transition from a high to a low activity level). However, this
possibility cannot be excluded when considering individual sources. Indeed,
the properties of 3C~028 fit nicely with this scheme: it has a classical FR~II
morphology, with well-defined twin jets linking the host to a pair of compact
hot spots, only lacking a radio core \citep{feretti84}. No continuum nuclear
emission is seen in the optical, infrared or X-ray bands
\citep{chiaberge:ccc,baldi10,balmaverde06a}. This appears to be an indication
that the drop in nuclear activity occurred very recently, which is confirmed
by its emission line luminosities: $L_{\rm H\alpha}$ is still within the range
of active RG, while $L_{\rm{ [O~III]}}$ is only reduced by a factor of
$\lesssim$ 10 (see Fig. \ref{lo3pcore}). This is what is expected if the
elapsed time since the change in the activity level is larger than the
radio-core cooling time and of the same order of $\tau_{\rm NLR}$. In this
source the \oiii\ might be still dominated by fossil emission.

However, this interpretation fails in the case of 3C~314.1. This source has a
relaxed double morphology \citep{leahy91}, which lacks of hot spots. This is
an indication that the switch-off of the jets in this galaxy occurred at least
$\sim0.7 \times 10^6$ years ago, which is the light travel time to the edge of
its 250 kpc radio source, a time interval much larger than the decay time of
the emission-lines considered, even including the geometric effects. No
significant fossil-line emission is expected after such a long time interval.
Furthermore (unlike the case of 3C~348 which we will discuss in more detail
later) no nuclear emission is observed in 3C~314.1 in the optical, 
infrared or radio bands.

These results suggest that in this source the line emission originates from
other emission mechanisms, which are qunrelated to nuclear photoionization. We
envisage two possibilities. In the first, the ionizing photon field is
produced by an evolved stellar population. \citet{stasinska08} show that pAGB
stars produce emission-lines whose ratios can mimic those observed in active
galaxies, including the low \oiii/\Hb\ ratios measured in relics. The analysis
of a large sample of SDSS spectra \citep{capetti11b,capetti11c} shows that in
this case the lines are closely linked to the stellar continuum, with a \oiii\
equivalent width , EW(\oiii), strongly clustered around 0.75 \AA. While for
3C~028 and 3C~348 the EW(\oiii) value is substantially larger ($\sim$5 \AA\
and $\sim$8 \AA, respectively), in 3C~314.1 we measure EW(\oiii) $\sim$1 \AA,
which is compatible with a stellar origin.

Radiative shocks, naturally expected in RGs, are also able to produce emission-lines with ratios similar to those observed in the relics \citep{allen08}. The
crossing time of such ``relic shocks'' across the NLR is sufficiently long
($\sim$$10^8$ years for a velocity of 100 \kms) to remain energetically
important for a long time after the drop in nuclear activity and the
consequent end of the input of energy from the jets.

The properties of 3C~348 suggest that this object underwent a transition from
a powerful RG to a lower power source. \citet{gizani03} proposed that 3C~348
may be a restarted source based on its peculiar radio properties: two jets
emerge from the core and propagate within a double-lobed radio structure which
is characterized by a very steep spectral index ($\alpha \sim$1.2) and lacks
any compact hot spots. Both jets show arc-like features, an indication of
episodic activity. The total radio luminosity at 178 MHz is L$_{178} = 2.2
\times 10^{35}$ \ergsHz, the second most luminous source in the z$<$0.3 3CR
sample. The multiwavelength nuclear properties of this source show instead a
strong similarity with those of FR~I.  The representative point of 3C~348
falls onto the correlation defined by FR~I in the plane comparing their
optical and radio emission \citep{chiaberge:ccc}. Also its \oiii\ luminosity
is similar to those of FR~I at the same level of core power (see
Fig. \ref{lo3pcore}). Although not detected in Chandra images, the upper limit
to its nuclear X-ray luminosity is consistent with the link between radio and
X-ray luminosities found for FR~I by \citet{balmaverde06core}. These results
suggest that the central engine of 3C~348 is now in a state characteristic of
FR~I. The current activity level might be a low state phase between multiple
outbursts, as its radio morphology suggests. The only difference between
3C~348 and the FR~Is (leaving aside the radio morphology and luminosity) is
the lower [O III]/\Hb\ ratio. The \Hb\ excess cannot be explained as a relic
of the previous phase of higher activity (given the $\sim$100 kpc size of its
FR~I radio structure) but possibly to shocks, similar to the case of 3C~314.1.

\section{Summary}
\label{summary}

We studied the properties of the ELEGs, the new spectroscopic class of
radio-loud AGN characterized by an extremely low \oiii/\Hb\ ratio
($\sim$0.5). They also show a deficit of \oiii\ emission and of radio-core
power with respect to RGs of similar radio luminosity. We interpret these
objects as relic AGN, i.e., galaxies that underwent a large drop in nuclear
activity. In the 3CR sample there are seven additional galaxies with a low
\oiii\ luminosity for their radio power that we identified as candidate
relics, but which were hitherto not classified spectroscopically.

We presented deeper observations for five candidate relics.  The newly measured
line ratios locate three galaxies in the optical spectroscopic diagnostic diagrams
outside the region characteristic of ELEGs, while they are insufficient to
establish the nature of the remaining two.  Therefore, we cannot confirm any
of the candidates as new ELEGs.

We also obtained further observations of the confirmed three ELEGs in order to
estimate their NLR gas density, a measurement that was not possible from the
survey spectra. We obtained fairly accurate measurements in the range $n_e =
160 - 300$ cm$^{-3}$, enabling us to estimate the emission-lines decay time
which results in only $t_d \sim$6-12 years for the \oiii.

We then explored the spectral evolution after the AGN ``death''.  As viewed by
a distant observer, the NLR can be considered divided into two regions bounded
by a parabolic iso-delay surface that corresponds to the active and fossil
states of the AGN. The thickness of the intermediate layer is much smaller
than the NLR, being only $t_d{\rm [O~III]} \times c \sim$2-4 pc. Thus, the NLR
temporal evolution is dominated by geometric effects. Indeed, the observed
spectrum is the superposition of the active and fossil regions, while
the contribution of the intermediate layer is negligible. The observed line
ratios are set by the relative amount of line emission in the two
regions. Thus, the characteristic timescale is the light travel time across
the NLR, $\tau_{\rm NLR} \sim 10{^4}$ years, and not the line-decay times.

The expected number of objects caught in the transition phase should then be
extremely small (of the order of $\tau_{\rm NLR}/\tau_{\rm RG} << 10^{-2} $),
in contrast to the fraction of ELEGs of a few percent found in the 3CR
sample. Furthermore, in two ELEGs the large-scale radio structure already
shows substantial changes with respect to the classical FR~II on a scale of
hundreds of kpc, indicating that the drop in the activity level occurred at
least $\sim$$10^6$ years ago.  No fossil line emission is expected in these
sources. Thus our results confirm that while relic RGs can be associated with
AGN that experienced a large drop in activity, the presence of
additional sources of ionization must be considered to understand their
spectroscopic properties.

Most likely, the class of relic RGs is composed of a mixed bag of objects in
different stages of their evolution. 3C~028 has a classical FR~II morphology,
with well defined jets and compact hot spots, but lacking any nuclear
emission. The deficit of line emission with respect to RGs of similar radio
power is modest, less than a factor of 10 even in \oiii. This suggests that
the drop in nuclear activity occurred very recently ($\lesssim 10^4$ years
ago) and that the emission-lines are still dominated by fossil emission.

In 3C~314.1 no nuclear emission is seen in the optical or radio bands. The
lack of hot spots in its radio source (extended over $\sim$250 kpc) indicates
that the AGN switch-off occurred at least $\sim$$10^6$ years ago. No fossil line
emission can be present. We suggest two alternatives in which the lines are
powered by i) shocks, whose crossing time of the NLR is sufficiently long,
$\sim$$10^8$ years, to be still active even in these late phases of evolution,
or ii) the ionizing continuum of an evolved stellar population, an
interpretation supported by the low EW of the emission-lines.

Finally, 3C~348 appears to be a transient from a powerful RG to a low
luminosity FR~I radio source. Considering the size of the FR~I radio structure
embedded in the radio relic ($\sim$100 kpc), the transition must have occurred
at least $\gtrsim 3\times 10^5$ years ago. 3C~348 shares the nuclear
properties of FR~Is from the point of view of the emission lines and the
radio and optical continuum. The \Hb\ excess, which causes the anomalous low
value of the \oiii/\Hb\ ratio, can be ascribed to relic shocks from its
past, high-power phase.

\acknowledgements
We wish to honor the memory of our great friend and colleague
David Axon.

\bibliographystyle{aa}

\end{document}